\documentclass[aps,twocolumn,prl,amsmath,amssymb,floatfix,showpacs]{revtex4}

\newcommand{\op}[1]{\hat{#1}}
\newcommand{\opdag}[1]{\hat{#1}^{\dag}}

\usepackage{braket}
\usepackage{graphicx}
\usepackage{subfigure}
\usepackage{ulem}
\usepackage{natbib}

\begin{document}

\title{Generalized rotating-wave approximation for arbitrarily large coupling}

\author{E. K. Irish}
\email{e.irish@qub.ac.uk}
\affiliation{School of Mathematics and Physics, Queen's University Belfast, Belfast BT7 1NN, UK}
\affiliation{Department of Physics and Astronomy, University of Rochester, Rochester, New York 14627, USA}

\date{\today}

\begin{abstract}
A generalized version of the rotating-wave approximation for the single-mode spin-boson Hamiltonian is presented. It is shown that performing a simple change of basis prior to eliminating the off-resonant terms results in a significantly more accurate expression for the energy levels of the system. The generalized approximation works for all values of the coupling strength and for a wide range of detuning values, and may find applications in solid-state experiments.
\end{abstract}

\maketitle

One of the simplest and most ubiquitous models in quantum physics is the single-mode spin-boson model, consisting of a two-level system coupled to a quantum harmonic oscillator. In quantum optics it describes an atom coupled to an electromagnetic field mode~\cite{Jaynes:1963,Shore:1993}; in condensed matter physics it lies at the heart of the Holstein model for electrons coupled to phonon modes of a crystal lattice~\cite{Holstein:1959a}. More recently, implementations of this model have been achieved in superconductor \cite{Chiorescu:2004,Wallraff:2004,Schuster:2007} and semiconductor \cite{Hennessy:2007} systems. Still other proposals have involved mechanical oscillators~\cite{Irish:2003,Schwab:2005}. Although the model itself is quite simple, it displays a rich variety of behaviors, encapsulating many of the unique aspects of quantum theory. 

The model Hamiltonian may be written as~\footnote{The notation used here is based on that commonly used for superconducting systems, which differs from the typical quantum optics notation by a rotation on the two-level system. Also, for simplicity, $\hbar$ is taken equal to $1$.}
\begin{equation}\label{H_tot}
H = \omega_0 \hat{a}^{\dag} \hat{a} +\tfrac{1}{2} \Omega \hat{\sigma}_x + \lambda \hat{\sigma}_z (\hat{a}^{\dag} + \hat{a}) .
\end{equation}
Despite decades of study, an analytical solution to this equation has not yet been found. A number of approximations have been developed, each tailored to a particular range of parameters. In quantum optics, one of the most useful approximations is the rotating-wave approximation (RWA), which is based on the assumption of near-resonance and relatively weak coupling between the two systems~\cite{Jaynes:1963,Shore:1993}. 

A generalization of the RWA that extends the range of validity to arbitrarily large coupling strengths is presented in this paper. The only difference from the ordinary RWA is that a change of basis is performed prior to carrying out the approximation. For the case of exact resonance ($\Omega = \omega_0$), the energy levels given here were first found by Amniat-Talab \textit{et al.}~\citep{Amniat-Talab:2005}. However, their derivation involved a complicated method of quantum averaging and resonant transformations. The derivation presented here is not restricted to exact resonance and the resulting approximation works remarkably well for large detuning. Moreover, in this form the simplicity of the approximation and its close connection to the standard RWA are emphasized.

To begin with, a brief review of the standard RWA is given in order to establish the arguments used in deriving the generalized approximation. The first step is to rewrite Eq.~\eqref{H_tot} in the form 
\begin{equation}\label{jcmham1}
\begin{split}
H &=  \omega_0 \opdag{a} \op{a} + \tfrac{1}{2}  \Omega \op{\sigma}_x +  \lambda (\op{\sigma}_- \opdag{a} + \op{\sigma}_+ \op{a} +  \op{\sigma}_+ \opdag{a} + \op{\sigma}_- \op{a})
\end{split}
\end{equation}
where $\op{\sigma}_{\pm} = \tfrac{1}{2} (\op{\sigma}_z \mp i \op{\sigma}_y)$ 
are the raising and lowering operators in the basis of $\op{\sigma}_x$. Alternatively, the Hamiltonian may be written in matrix form in the basis $\ket{\pm x, N}$ (where $N = 0,1,2,\dots$), which is the eigenbasis of the noninteracting Hamiltonian $H_0 =  \omega_0 \opdag{a} \op{a} + \tfrac{1}{2} \Omega \op{\sigma}_x$:
\begin{equation}\label{jcmmatrix1}
H = 
\begin{pmatrix} 
E_{-,0}^{(0)} & 0 & 0 &  \lambda & 0 & 0 & \dots \\
0 & E_{+,0}^{(0)} &  \lambda & 0 & 0 & 0 & \dots \\
0 &  \lambda & E_{-,1}^{(0)} & 0 & 0 & \sqrt{2}  \lambda & \dots \\
 \lambda & 0 & 0 & E_{+,1}^{(0)} & \sqrt{2}  \lambda & 0 & \dots \\ 
0 & 0 & 0 & \sqrt{2}  \lambda & E_{-,2}^{(0)} & 0 & \dots \\
0 & 0 & \sqrt{2}  \lambda & 0 & 0 & E_{+,2}^{(0)} & \dots \\
\vdots & \vdots & \vdots & \vdots & \vdots & \vdots &  \ddots
\end{pmatrix} 
\end{equation}
where $E_{\pm,N}^{(0)} = N  \omega_0 \pm \tfrac{1}{2} \Omega$ and the order of the columns and rows is $\ket{-x,0}, \ket{+x,0}, \ket{-x,1}, \ket{+x,1}, \ldots$.

Consider the case of near-resonance ($\omega_0 \approx \Omega$) and weak coupling ($\lambda \ll \omega_0, \Omega$). The interaction term $ \op{\sigma}_- \opdag{a} + \op{\sigma}_+ \op{a}$ couples the states $\ket{+x,N}$ and $\ket{-x,N+1}$, which have nearly equal energies in the absence of the interaction. On the other hand, the term $ \op{\sigma}_+ \opdag{a} + \op{\sigma}_- \op{a}$ couples the off-resonant states $\ket{-x,N}$ and $\ket{+x,N+1}$. In this sense the first term is ``energy conserving,'' while the second is not. The rotating-wave approximation eliminates the non-energy-conserving terms. In matrix form this corresponds to removing the remote matrix elements. The Hamiltonian then becomes block diagonal and may be readily diagonalized.

Alternatively, the RWA Hamiltonian may be derived by moving to the interaction picture with respect to $H_0$. The Hamiltonian becomes
\begin{equation}
\begin{split}
H_1^I(t) &= \exp(i H_0 t) \lambda \hat{\sigma}_z (\hat{a}^{\dag} + \hat{a}) \exp(-i H_0 t) \\
&= \lambda (\op{\sigma}_- \opdag{a}e^{i (\omega_0 - \Omega) t} + \op{\sigma}_+ \op{a}e^{-i (\omega_0 - \Omega) t} \\
&\quad + \op{\sigma}_+ \opdag{a}e^{i (\omega_0 + \Omega) t} + \op{\sigma}_- \op{a}e^{-i (\omega_0 + \Omega) t}) .
\end{split}
\end{equation}
In the case of near resonance, $\omega_0 \approx \Omega$ and the first two terms vary slowly in time. The last two terms, however, vary rapidly and therefore average to zero over timescales on the order of $1/\omega_0$. Thus the 
last two terms may be neglected and, transforming back to the Schr{\"o}dinger picture, the coupling reduces to $ \lambda(\op{\sigma}_- \opdag{a} + \op{\sigma}_+ \op{a})$. These are exactly the ``energy conserving'' terms discussed in the matrix derivation. Indeed, the argument about energy conservation and the argument about timescales are completely equivalent. 

The next task is to establish a new set of basis states. These are the states obtained from an adiabatic approximation in the limit $\Omega \ll (\omega_0, \lambda)$. Several derivations of this approximation have been presented~\citep{Graham:1984a,Schweber:1967,Crisp:1992}. However, the matrix-based derivation given in Ref.~\citep{Irish:2005} is the most useful for the purposes of this paper and is briefly summarized here.

The adiabatic approximation is most readily carried out in the basis obtained by setting $\Omega = 0$ in Eq.~\eqref{H_tot}:
\begin{subequations}\label{displ_osc }
\begin{gather}
\ket{\pm z, N_{\pm}} \equiv \ket{\pm z} \otimes e^{\mp (\lambda/\omega_0)(\opdag{a} - \op{a})} \ket{N} , \label{displ_osc_states} \\
E_N =  \omega_0 (N - \lambda^2/\omega_0^2) .\label{displ_osc_energies}
\end{gather}
\end{subequations}
The qubit states $\ket{\pm z}$ are eigenstates of $\op{\sigma}_z$ and the oscillator states $\ket{N_{\pm}}$ are position-displaced Fock states. Note that $\ket{+z, N_{+}}$ and $\ket{-z, N_{-}}$ are degenerate in energy.

The spin term $\tfrac{1}{2} \Omega \hat{\sigma}_x$ couples the basis states given in Eq.~\eqref{displ_osc_states}. Within the adiabatic approximation, only the coupling between states with the same value of $N$ is considered. In matrix form, this corresponds to reducing the matrix to a block diagonal form, where the blocks are given by
\begin{equation}\label{ad_approx_diag_mat}
\begin{pmatrix}
E_N & \tfrac{1}{2}  \Omega \braket{N_- | N_+} \\
\tfrac{1}{2}  \Omega \braket{N_- | N_+} & E_N
\end{pmatrix} .
\end{equation}
The expression $\braket{N_- | N_+}$ is simply the overlap of the two position-displaced Fock states, given by ($M \le N$)
\begin{equation}\label{overlap}
\braket{M_- | N_+ } = e^{-2 \lambda^2 / \omega_0^2} \negthinspace \left(\frac{2 \lambda}{\omega_0}\right)^{N-M} \negthickspace \negmedspace \sqrt{\frac{M!}{N!}}L_{M}^{N-M} \negmedspace \left(\frac{4 \lambda^2}{\omega_0^2}\right) .
\end{equation}
The $2\times2$ matrix of Eq.~\eqref{ad_approx_diag_mat} has the eigenstates and energies
\begin{subequations}\label{eigen1}
\begin{gather}
\ket{\Psi_{\pm, N}}= \tfrac{1}{\sqrt{2}}(\ket{+z, N_+} \pm \ket{-z, N_-}) , \label{states1} \\
E_{\pm, N} = \pm \tfrac{1}{2}  \Omega \langle N_- \vert N_+ \rangle + E_N . \label{energy1} 
\end{gather}
\end{subequations} 
An analysis of the adiabatic approximation and its consequences may be found in Ref.~\citep{Irish:2005}. 

The derivation of the generalized rotating-wave approximation (GRWA) is now quite straightforward. The Hamiltonian is rewritten in the basis of the adiabatic eigenstates $\ket{\Psi_{\pm, N}}$. Then the argument about energy conservation that led to the RWA is applied in the new basis and the approximate energy levels are calculated. 

When written in the basis of the states $\ket{\Psi_{-,0}}, \ket{\Psi_{+,0}}, \ket{\Psi_{-,1}}, \ket{\Psi_{+,1}}, \dots$, Eq.~\eqref{H_tot} becomes
\begin{equation}\label{gen_matrix1}
H =
\begin{pmatrix}
E_{-, 0} & 0 & 0 & -\tfrac{1}{2}\Omega^{\prime}_{0,1} & \tfrac{1}{2}\Omega^{\prime}_{0,2} & \dots \\
0 & E_{+,0} & \tfrac{1}{2}\Omega^{\prime}_{0,1} & 0 & 0  & \dots \\
0 & \tfrac{1}{2}\Omega^{\prime}_{0,1} & E_{-,1} & 0 & 0 & \dots \\
-\tfrac{1}{2}\Omega^{\prime}_{0,1} & 0 & 0 & E_{+,1} & \tfrac{1}{2}\Omega^{\prime}_{1,2} & \dots \\ 
\tfrac{1}{2}\Omega^{\prime}_{0,2} & 0 & 0 & \tfrac{1}{2}\Omega^{\prime}_{1,2} & E_{-,2} & \dots \\
\vdots & \vdots & \vdots & \vdots & \vdots & \ddots
\end{pmatrix} ,
\end{equation}
where $\Omega^{\prime}_{M,N} \equiv \Omega \braket{M_-|N_+}$. The form of this matrix closely resembles that of Eq.~\eqref{jcmmatrix1} with additional remote matrix elements. As before, the approximation consists of neglecting the remote matrix elements, reducing the matrix to a $2 \times 2$ block diagonal form. 

Although it is not immediately evident from the matrix form, the terms retained in this approximation correspond to energy-conserving one-particle transitions, just as in the ordinary RWA. This is most easily illustrated in the interaction picture. First the change of basis from $\ket{\mp x, N}$ to $\ket{\Psi_{\mp, N}}$ is carried out by a unitary transformation with the operator
\begin{equation}
\op{D}(\tfrac{\lambda}{\omega_0} \op{\sigma}_z) =  \exp[-\tfrac{\lambda}{\omega_0} \op{\sigma}_z (\opdag{a} - \op{a})] ,
\end{equation}
which is a spin-dependent position displacement operator. Applying this transformation to Eq.~\eqref{H_tot} results in the transformed Hamiltonian
\begin{align}
\tilde{H} &= \opdag{D} H \op{D} \nonumber \\
&=  \omega_0 \opdag{a} \op{a} + \tfrac{1}{2} \Omega \op{\sigma}_x \exp[-\tfrac{2 \lambda}{\omega_0} \op{\sigma}_z (\opdag{a} - \op{a})] \\
&=  \omega_0 \opdag{a} \op{a} + \tfrac{1}{2} \Omega \op{\sigma}_x + \tilde{H}_{1,x} + \tilde{H}_{1,y},
\end{align}
where the functions $\tilde{H}_{1,x}$ and $\tilde{H}_{1,y}$ are defined as
\begin{align}
\tilde{H}_{1,x} &=  \tfrac{1}{2} \Omega \op{\sigma}_x \left[2(\tfrac{\lambda}{\omega_0})^2 (\opdag{a} - \op{a})^2 + \tfrac{2}{3}(\tfrac{\lambda}{\omega_0})^4 (\opdag{a} - \op{a})^4 + \dots \right] \\
\tilde{H}_{1,y} &=  \tfrac{i}{2} \Omega \op{\sigma}_y \left[2(\tfrac{\lambda}{\omega_0}) (\opdag{a} - \op{a}) + \tfrac{4}{3}(\tfrac{\lambda}{\omega_0})^3 (\opdag{a} - \op{a})^3 + \dots \right] .
\end{align} 
The next step is to move to the interaction picture with respect to $\tilde{H}_0 =  \omega_0 \opdag{a} \op{a} + \tfrac{1}{2}  \Omega \op{\sigma}_x$. Let us examine $\tilde{H}_{1,x}$ first. Since $\op{\sigma}_x$ commutes with the rotation operator $\op{U} = \exp(i \tilde{H}_0 t)$, the rotation affects only the oscillator operators. Take the first term of $\tilde{H}_{1,x}$ as an example. In the interaction picture the operators become
\begin{equation}
\op{\sigma}_x (-2\opdag{a} \op{a} - 1 + \op{a}^{\dag 2} e^{2 i \omega_0 t} + \op{a}^2 e^{-2 i \omega_0 t}) .
\end{equation}
The time-independent terms contain powers of the number operator $\opdag{a} \op{a}$ and correspond to transitions that result in zero net excitation of the oscillator. They are diagonal in the basis $\ket{\Psi_{\pm,N}}$ and modify the spin frequency $\Omega$, resulting in the term $\pm \tfrac{1}{2} \Omega \braket{N_- | N_+}$ that appears in $E_{\pm,N}$. The higher-order terms, which have a rapid time dependence, produce remote matrix elements such as the two-excitation term $\braket{\Psi_{-,0} | H | \Psi_{-,2}}$ and are neglected within the GRWA.

Next consider $\tilde{H}_{1,y}$. Taking $i \op{\sigma}_y = \tfrac{1}{2}(\op{\sigma}_- - \op{\sigma}_+)$, the first term is proportional to
\begin{equation}
\begin{split}
\bigl[&\opdag{a} \op{\sigma}_- e^{i(\omega_0 - \Omega) t} + \op{a} \op{\sigma}_+ e^{-i(\omega_0 - \Omega) t} \\
&- \opdag{a} \op{\sigma}_+ e^{i(\omega_0 + \Omega) t} - \op{a} \op{\sigma}_+ e^{-i(\omega_0 + \Omega) t}\bigr] .
\end{split}
\end{equation}
The second term of $\tilde{H}_{1,y}$ is slightly more complicated. When $(\opdag{a} - \op{a})^3$ is expanded and put into normal order, this term is given in the interaction picture by
\begin{equation}\label{thirdorderterm}
\begin{split}
\bigl[&-3 \op{\sigma}_- \opdag{a}(\opdag{a} \op{a} + 1) e^{i (\omega_0 - \Omega) t} + 3 \op{\sigma}_+ (\opdag{a} \op{a} + 1) \op{a }e^{-i (\omega_0 - \Omega) t} \\
&-3 \op{\sigma}_+ \opdag{a}(\opdag{a} \op{a} + 1) e^{i (\omega_0 + \Omega) t} + 3 \op{\sigma}_- (\opdag{a} \op{a} + 1) \op{a }e^{-i (\omega_0 + \Omega) t} \\
&+ \op{\sigma}_- \op{a}^{\dag 3} e^{i (3 \omega_0 - \Omega) t} - \op{\sigma}_+ \op{a}^3 e^{-i (3 \omega_0 - \Omega) t} \\
&+ \op{\sigma}_+ \op{a}^{\dag 3} e^{i (3 \omega_0 + \Omega) t} - \op{\sigma}_- \op{a}^3 e^{-i (3 \omega_0 + \Omega) t}\bigr] .
\end{split}
\end{equation}
The first two terms create energy-conserving transitions involving a single excitation. They produce the matrix elements $\braket{\Psi_{+,N} | H | \Psi_{-, N+1}}$ and $\braket{\Psi_{-,N+1} | H | \Psi_{+,N}}$ that appear immediately off the diagonal in Eq.~\eqref{gen_matrix1}. The next two terms correspond to energy non-conserving single-excitation transitions and produce the remote matrix elements $\braket{\Psi_{-,N} | H | \Psi_{+, N+1}}$ and $\braket{\Psi_{+,N+1} | H | \Psi_{-,N}}$. The last four terms involve a net change of three excitations and produce remote matrix elements. Only the first two terms of Eq.~\eqref{thirdorderterm} have slow time dependence when $\Omega \approx \omega_0$.
 
Finally, the GRWA is carried out by keeping only the ``energy-conserving'' one-excitation terms. The other one-excitation terms as well as terms involving higher numbers of quanta are discarded. When all powers of $\lambda/\omega_0$ are taken into account, $\tilde{H}_{1,y}$ reduces to a coupling term of the form
\begin{equation}\label{gencoupling}
\Omega (\tfrac{\lambda}{\omega_0})[\op{\sigma}_- \opdag{a} f(\opdag{a} \op{a}) + \op{\sigma}_+ f^*(\opdag{a} \op{a}) \op{a}] ,
\end{equation}
where the function $f(\opdag{a} \op{a})$ is too complicated to display here. Equation~\eqref{gencoupling} is a generalization of the energy-conserving term $ \lambda (\op{\sigma}_- \opdag{a} + \op{\sigma}_+ \op{a})$ in the usual RWA Hamiltonian~\footnote{Interestingly enough, Eq.~\eqref{gencoupling} reduces to the standard RWA coupling in the limit $\lambda/\omega_0 \ll 1$ when $\Omega = \omega_0$.}. 

\begin{figure} 
\begin{center}
\includegraphics[scale=1]{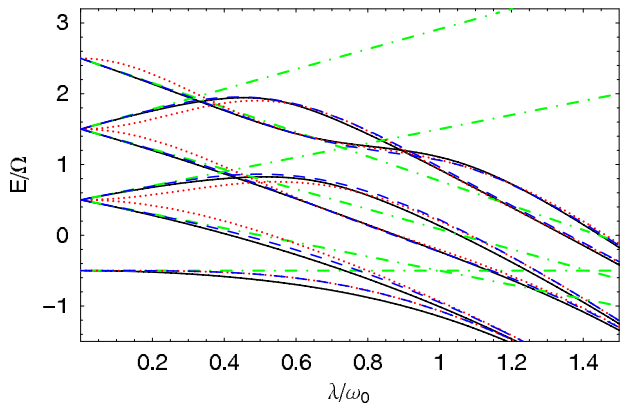}
\caption{\label{fig:on_resonance} Comparison of the RWA (dot-dashed), adiabatic approximation (dotted), and GRWA (dashed) with numerically-determined energy levels (solid) in the resonance case, $\omega_0 = \Omega$. }
\end{center}
\end{figure}

Returning to the matrix picture, the GRWA ground state is uncoupled from all the other states (just as in the RWA), so the ground state energy is given by $E_{-,0}$. The remainder of the matrix takes the familiar $2 \times 2$ block-diagonal structure with blocks of the form
\begin{equation}\label{gen_rwa}
\begin{pmatrix}
E_{+,N-1} & \tfrac{1}{2} \Omega^{\prime}_{N-1,N}\\
\tfrac{1}{2} \Omega^{\prime}_{N-1,N} & E_{-,N}
\end{pmatrix} .
\end{equation}
Solving for the eigenvalues of the blocks yields the GRWA energies:
\begin{widetext}
\begin{equation}\label{gen_rwa_energies}
\begin{split}
E_{\pm, N}^{\text{GRWA}} &= (N + \tfrac{1}{2}) \omega_0 - \frac{\lambda^2}{\omega_0} + \frac{\Omega}{4} e^{-2 \lambda^2/\omega_0^2} [L_N(4 \lambda^2/\omega_0^2) - L_{N+1}(4 \lambda^2/\omega_0^2)] \\
&\quad \pm \biggl( \Bigl\lbrace  \tfrac{1}{2} \omega_0 - \tfrac{1}{4} \Omega e^{-2 \lambda^2/\omega_0^2} \left[ L_N(4 \lambda^2/\omega_0^2) + L_{N+1}(4 \lambda^2/\omega_0^2) \right] \Bigr\rbrace^2 
+ \frac{\lambda^2 \Omega^2}{\omega_0^2(N+1)} e^{-4 \lambda^2/\omega_0^2} \left[ L_N^1(4 \lambda^2/\omega_0^2) \right]^2 \biggr)^{1/2} .
\end{split}
\end{equation}
\end{widetext}
The energy levels from the RWA, the adiabatic approximation, and the GRWA are plotted in Fig.~\ref{fig:on_resonance}. For comparison purposes, the energy levels obtained from a numerical solution of Eq.~\eqref{H_tot} are also shown. The RWA reproduces the correct limiting behavior as $\lambda/\omega_0 \to 0$, but breaks down near the point where the paired levels first cross. On the other hand, the adiabatic approximation diverges from the numerical solution at small values of $\lambda/\omega_0$, but captures the behavior beyond the first crossing point very well. The GRWA combines the behavior of the adiabatic approximation at large values of $\lambda$ with the accuracy of the RWA at small values, providing an excellent approximation to the actual energies of the system over the full range of coupling strengths shown.

\begin{figure}
\begin{center}
\includegraphics[scale=1.0]{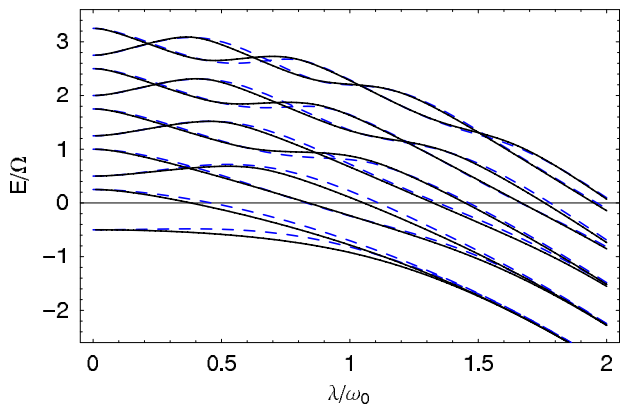}
\caption{\label{fig:off_resonance} GRWA energy levels (dashed lines) compared with numerically-determined energies (solid lines) in the off-resonance case, $\omega_0 = 0.75 \Omega$.}
\end{center}
\end{figure}

Remarkably, the GRWA works reasonably well even for large detunings with $\omega_0 < \Omega$. As an example, the case $\omega_0 = 0.75 \Omega$ is illustrated in Fig.~\ref{fig:off_resonance}. The maximum error in the energy is less than $0.2 \omega_0$ for the ground state and decreases for higher energy levels. The qualitative agreement between the GRWA and the exact solution remains fairly good even down to $\omega_0 = 0.5 \Omega$. Considering that the RWA requires small detuning and the adiabatic approximation is derived under the assumption that $\omega_0 \gg \Omega$, the GRWA is surprisingly robust in this parameter regime. 

Why does the GRWA work so well? It seems counterintuitive that a simple change of basis for the RWA should result in such an improvement. One explanation comes from examining a fundamental similarity between the RWA and the adiabatic approximation: both involve calculating the energy splitting due to an interaction between two otherwise degenerate basis states. In the RWA, the degeneracy or resonance occurs at the single point ($\Omega=\omega_0$, $\lambda = 0$). The adiabatic approximation, on the other hand, treats the resonance at $\Omega = 0$, which occurs for all values of $\lambda$. This interpretation accounts for the fact that the RWA only works for small $\lambda$ as well as for the accuracy of the adiabatic approximation at all values of $\lambda$ when $\Omega \ll \omega_0$. The adiabatic approximation breaks down at small $\lambda$ when $\Omega = \omega_0$ precisely because it does not account for the zero-coupling resonance~\citep{Amniat-Talab:2005}. However, the GRWA takes into account both the resonance at $\Omega=0$ and the point-like resonance at ($\Omega = \omega_0, \lambda = 0$), which yields a very accurate energy spectrum.

One reason the standard RWA has remained so prevalent in quantum optics is that
the conditions of near-resonance and weak coupling are naturally satisfied in atomic cavity quantum electrodynamics (CQED) experiments~\cite{Hood:2000,Raimond:2001}. The RWA works extremely well for such systems. However, superconducting CQED-like systems are capable of much higher coupling strengths, even at large detunings, and are already nearing the limits of validity of the RWA~\cite{Chiorescu:2004,Wallraff:2004,Schuster:2007}. The generalized approximation presented in this paper provides an excellent treatment for the very strong coupling limit that these experiments are expected to achieve, while maintaining strong links to the familiar language and techniques of quantum optics. Thus the GRWA may prove useful as experiments continue to expand the accessible parameter regime in this important and still fascinating model.

\begin{acknowledgments}
I would like to thank N.~P. Bigelow, J. Gea-Banacloche, M.~S. Kim, and M. Paternostro for their helpful comments and encouragement. Support from the National Physical Sciences Consortium and the EPSRC is gratefully acknowledged.
\end{acknowledgments}


\begin{thebibliography}{16}
\expandafter\ifx\csname natexlab\endcsname\relax\def\natexlab#1{#1}\fi
\expandafter\ifx\csname bibnamefont\endcsname\relax
  \def\bibnamefont#1{#1}\fi
\expandafter\ifx\csname bibfnamefont\endcsname\relax
  \def\bibfnamefont#1{#1}\fi
\expandafter\ifx\csname citenamefont\endcsname\relax
  \def\citenamefont#1{#1}\fi
\expandafter\ifx\csname url\endcsname\relax
  \def\url#1{\texttt{#1}}\fi
\expandafter\ifx\csname urlprefix\endcsname\relax\def\urlprefix{URL }\fi
\providecommand{\bibinfo}[2]{#2}
\providecommand{\eprint}[2][]{\url{#2}}

\bibitem[{\citenamefont{Jaynes and Cummings}(1963)}]{Jaynes:1963}
\bibinfo{author}{\bibfnamefont{E.~T.} \bibnamefont{Jaynes}} \bibnamefont{and}
  \bibinfo{author}{\bibfnamefont{F.~W.} \bibnamefont{Cummings}},
  \bibinfo{journal}{Proc. IEEE} \textbf{\bibinfo{volume}{51}},
  \bibinfo{pages}{89} (\bibinfo{year}{1963}).

\bibitem[{\citenamefont{Shore and Knight}(1993)}]{Shore:1993}
\bibinfo{author}{\bibfnamefont{B.~W.} \bibnamefont{Shore}} \bibnamefont{and}
  \bibinfo{author}{\bibfnamefont{P.~L.} \bibnamefont{Knight}},
  \bibinfo{journal}{J. Mod. Optics} \textbf{\bibinfo{volume}{40}},
  \bibinfo{pages}{1195} (\bibinfo{year}{1993}).

\bibitem[{\citenamefont{Holstein}(1959)}]{Holstein:1959a}
\bibinfo{author}{\bibfnamefont{T.}~\bibnamefont{Holstein}},
  \bibinfo{journal}{Ann. Phys. (N.Y.)} \textbf{\bibinfo{volume}{8}},
  \bibinfo{pages}{325} (\bibinfo{year}{1959}).

\bibitem[{\citenamefont{Chiorescu et~al.}(2004)\citenamefont{Chiorescu, Bertet,
  Semba, Nakamura, Harmans, and Mooij}}]{Chiorescu:2004}
\bibinfo{author}{\bibfnamefont{I.}~\bibnamefont{Chiorescu}},
  \bibinfo{author}{\bibfnamefont{P.}~\bibnamefont{Bertet}},
  \bibinfo{author}{\bibfnamefont{K.}~\bibnamefont{Semba}},
  \bibinfo{author}{\bibfnamefont{Y.}~\bibnamefont{Nakamura}},
  \bibinfo{author}{\bibfnamefont{C.~J. P.~M.} \bibnamefont{Harmans}},
  \bibnamefont{and} \bibinfo{author}{\bibfnamefont{J.~E.} \bibnamefont{Mooij}},
  \bibinfo{journal}{Nature (London)} \textbf{\bibinfo{volume}{431}},
  \bibinfo{pages}{159} (\bibinfo{year}{2004}).

\bibitem[{\citenamefont{Wallraff et~al.}(2004)\citenamefont{Wallraff, Schuster,
  Blais, Frunzio, Huang, Majer, Kumar, Girvin, and Schoelkopf}}]{Wallraff:2004}
\bibinfo{author}{\bibfnamefont{A.}~\bibnamefont{Wallraff}},
  \bibinfo{author}{\bibfnamefont{D.~I.} \bibnamefont{Schuster}},
  \bibinfo{author}{\bibfnamefont{A.}~\bibnamefont{Blais}},
  \bibinfo{author}{\bibfnamefont{L.}~\bibnamefont{Frunzio}},
  \bibinfo{author}{\bibfnamefont{R.~S.} \bibnamefont{Huang}},
  \bibinfo{author}{\bibfnamefont{J.}~\bibnamefont{Majer}},
  \bibinfo{author}{\bibfnamefont{S.}~\bibnamefont{Kumar}},
  \bibinfo{author}{\bibfnamefont{S.~M.} \bibnamefont{Girvin}},
  \bibnamefont{and} \bibinfo{author}{\bibfnamefont{R.~J.}
  \bibnamefont{Schoelkopf}}, \bibinfo{journal}{Nature (London)}
  \textbf{\bibinfo{volume}{431}}, \bibinfo{pages}{162} (\bibinfo{year}{2004}).

\bibitem[{\citenamefont{Schuster et~al.}(2007)\citenamefont{Schuster, Houck,
  Schreier, Wallraff, Gambetta, Blais, Frunzio, Majer, Johnson, Devoret
  et~al.}}]{Schuster:2007}
\bibinfo{author}{\bibfnamefont{D.~I.} \bibnamefont{Schuster}},
  \bibinfo{author}{\bibfnamefont{A.~A.} \bibnamefont{Houck}},
  \bibinfo{author}{\bibfnamefont{J.~A.} \bibnamefont{Schreier}},
  \bibinfo{author}{\bibfnamefont{A.}~\bibnamefont{Wallraff}},
  \bibinfo{author}{\bibfnamefont{J.~M.} \bibnamefont{Gambetta}},
  \bibinfo{author}{\bibfnamefont{A.}~\bibnamefont{Blais}},
  \bibinfo{author}{\bibfnamefont{L.}~\bibnamefont{Frunzio}},
  \bibinfo{author}{\bibfnamefont{J.}~\bibnamefont{Majer}},
  \bibinfo{author}{\bibfnamefont{B.}~\bibnamefont{Johnson}},
  \bibinfo{author}{\bibfnamefont{M.~H.} \bibnamefont{Devoret}},
  \bibnamefont{et~al.}, \bibinfo{journal}{Nature (London)}
  \textbf{\bibinfo{volume}{445}}, \bibinfo{pages}{515} (\bibinfo{year}{2007}).

\bibitem[{\citenamefont{Hennessy et~al.}(2007)\citenamefont{Hennessy, Badolato,
  Winger, Gerace, Atat{\"u}re, Gulde, F{\"a}lt, Hu, and Imamo{\u
  g}lu}}]{Hennessy:2007}
\bibinfo{author}{\bibfnamefont{K.}~\bibnamefont{Hennessy}},
  \bibinfo{author}{\bibfnamefont{A.}~\bibnamefont{Badolato}},
  \bibinfo{author}{\bibfnamefont{M.}~\bibnamefont{Winger}},
  \bibinfo{author}{\bibfnamefont{D.}~\bibnamefont{Gerace}},
  \bibinfo{author}{\bibfnamefont{M.}~\bibnamefont{Atat{\"u}re}},
  \bibinfo{author}{\bibfnamefont{S.}~\bibnamefont{Gulde}},
  \bibinfo{author}{\bibfnamefont{S.}~\bibnamefont{F{\"a}lt}},
  \bibinfo{author}{\bibfnamefont{E.~L.} \bibnamefont{Hu}}, \bibnamefont{and}
  \bibinfo{author}{\bibfnamefont{A.}~\bibnamefont{Imamo{\u g}lu}},
  \bibinfo{journal}{Nature (London)} \textbf{\bibinfo{volume}{445}},
  \bibinfo{pages}{896} (\bibinfo{year}{2007}).

\bibitem[{\citenamefont{Irish and Schwab}(2003)}]{Irish:2003}
\bibinfo{author}{\bibfnamefont{E.~K.} \bibnamefont{Irish}} \bibnamefont{and}
  \bibinfo{author}{\bibfnamefont{K.}~\bibnamefont{Schwab}},
  \bibinfo{journal}{Phys. Rev. B} \textbf{\bibinfo{volume}{68}},
  \bibinfo{pages}{155311} (\bibinfo{year}{2003}).

\bibitem[{\citenamefont{Schwab and Roukes}(2005)}]{Schwab:2005}
\bibinfo{author}{\bibfnamefont{K.~C.} \bibnamefont{Schwab}} \bibnamefont{and}
  \bibinfo{author}{\bibfnamefont{M.~L.} \bibnamefont{Roukes}},
  \bibinfo{journal}{Phys. Today} \textbf{\bibinfo{volume}{58}},
  \bibinfo{pages}{36} (\bibinfo{year}{2005}).

\bibitem[{\citenamefont{Amniat-Talab et~al.}(2005)\citenamefont{Amniat-Talab,
  Gu{\'e}rin, and Jauslin}}]{Amniat-Talab:2005}
\bibinfo{author}{\bibfnamefont{M.}~\bibnamefont{Amniat-Talab}},
  \bibinfo{author}{\bibfnamefont{S.}~\bibnamefont{Gu{\'e}rin}},
  \bibnamefont{and} \bibinfo{author}{\bibfnamefont{H.~R.}
  \bibnamefont{Jauslin}}, \bibinfo{journal}{J. Math. Phys.}
  \textbf{\bibinfo{volume}{46}}, \bibinfo{pages}{042311}
  (\bibinfo{year}{2005}).

\bibitem[{\citenamefont{Graham and H{\"o}hnerbach}(1984)}]{Graham:1984a}
\bibinfo{author}{\bibfnamefont{R.}~\bibnamefont{Graham}} \bibnamefont{and}
  \bibinfo{author}{\bibfnamefont{M.}~\bibnamefont{H{\"o}hnerbach}},
  \bibinfo{journal}{Z. Phys. B: Condens. Matter} \textbf{\bibinfo{volume}{57}},
  \bibinfo{pages}{233} (\bibinfo{year}{1984}).

\bibitem[{\citenamefont{Schweber}(1967)}]{Schweber:1967}
\bibinfo{author}{\bibfnamefont{S.}~\bibnamefont{Schweber}},
  \bibinfo{journal}{Ann. Phys. (N.Y.)} \textbf{\bibinfo{volume}{41}},
  \bibinfo{pages}{205} (\bibinfo{year}{1967}).

\bibitem[{\citenamefont{Crisp}(1992)}]{Crisp:1992}
\bibinfo{author}{\bibfnamefont{M.~D.} \bibnamefont{Crisp}},
  \bibinfo{journal}{Phys. Rev. A} \textbf{\bibinfo{volume}{46}},
  \bibinfo{pages}{4138} (\bibinfo{year}{1992}).

\bibitem[{\citenamefont{Irish et~al.}(2005)\citenamefont{Irish, Gea-Banacloche,
  Martin, and Schwab}}]{Irish:2005}
\bibinfo{author}{\bibfnamefont{E.~K.} \bibnamefont{Irish}},
  \bibinfo{author}{\bibfnamefont{J.}~\bibnamefont{Gea-Banacloche}},
  \bibinfo{author}{\bibfnamefont{I.}~\bibnamefont{Martin}}, \bibnamefont{and}
  \bibinfo{author}{\bibfnamefont{K.~C.} \bibnamefont{Schwab}},
  \bibinfo{journal}{Phys. Rev. B} \textbf{\bibinfo{volume}{72}},
  \bibinfo{pages}{195410} (\bibinfo{year}{2005}).

\bibitem[{\citenamefont{Hood et~al.}(2000)\citenamefont{Hood, Lynn, Doherty,
  Parkins, and Kimble}}]{Hood:2000}
\bibinfo{author}{\bibfnamefont{C.~J.} \bibnamefont{Hood}},
  \bibinfo{author}{\bibfnamefont{T.~W.} \bibnamefont{Lynn}},
  \bibinfo{author}{\bibfnamefont{A.~C.} \bibnamefont{Doherty}},
  \bibinfo{author}{\bibfnamefont{A.~S.} \bibnamefont{Parkins}},
  \bibnamefont{and} \bibinfo{author}{\bibfnamefont{H.~J.}
  \bibnamefont{Kimble}}, \bibinfo{journal}{Science}
  \textbf{\bibinfo{volume}{287}}, \bibinfo{pages}{1447} (\bibinfo{year}{2000}).

\bibitem[{\citenamefont{Raimond et~al.}(2001)\citenamefont{Raimond, Brune, and
  Haroche}}]{Raimond:2001}
\bibinfo{author}{\bibfnamefont{J.~M.} \bibnamefont{Raimond}},
  \bibinfo{author}{\bibfnamefont{M.}~\bibnamefont{Brune}}, \bibnamefont{and}
  \bibinfo{author}{\bibfnamefont{S.}~\bibnamefont{Haroche}},
  \bibinfo{journal}{Rev. Mod. Phys.} \textbf{\bibinfo{volume}{73}},
  \bibinfo{pages}{565} (\bibinfo{year}{2001}).

\end{thebibliography}
\end{document}